\documentclass[usenatbib]{mn2e}

\usepackage{epsfig}
\usepackage{mathrsfs,aas_macros}

\def\Lya{Ly$\alpha$~}

\def\HI{\hbox{H$\,\rm \scriptstyle I\ $}}
 
\def\HeI{\hbox{He$\,\rm \scriptstyle I\ $}}
\def\HeII{\hbox{He$\,\rm \scriptstyle II\ $}}
\def\HeIII{\hbox{He$\,\rm \scriptstyle III\ $}}

\title[IGM temperature measurements around $z\simeq 6$ quasars]{Improved measurements of the intergalactic medium temperature around quasars: possible evidence for the initial stages of \HeII reionisation at {\boldmath $z\simeq 6$}}

\author[J.S. Bolton et al.] {James S. Bolton$^{1}$, George
  D. Becker$^{2}$, Sudhir Raskutti$^{1}$, J. Stuart B. Wyithe$^{1}$,
  \newauthor Martin G. Haehnelt$^{2}$, \& Wallace
  L.W. Sargent$^{3}$\\ $^1$ School of Physics, University of
  Melbourne, Parkville, VIC 3010, Australia \\ $^2$ Kavli Institute
  for Cosmology and Institute of Astronomy, Madingley Road, Cambridge,
  CB3 0HA \\ $^3$ Palomar Observatory, California Institute of
  Technology, Pasadena, CA 91125, USA}

\begin{document}

\date{}

\maketitle

\label{firstpage}

\begin{abstract}

We present measurements of the intergalactic medium (IGM) temperature
within $\sim 5$ proper Mpc of seven luminous quasars at $z \simeq 6$.
The constraints are obtained from the Doppler widths of \Lya
absorption lines in the quasar near-zones and build upon our previous
measurement for the $z=6.02$ quasar SDSS J0818$+$1722.  The expanded
data set, combined with an improved treatment of systematic
uncertainties, yields an average temperature at the mean density of
$\log (T_{0}/\rm K) = 4.21 \pm^{0.03}_{0.03}$ ($\pm^{0.06}_{0.07}$) at 68
(95) per cent confidence for a flat prior distribution over $3.2 \leq
\log (T_{0}/\rm K) \leq 4.8$.  In comparison, temperatures measured
from the general IGM at $z\simeq 5$ are $\sim 0.3$ dex cooler,
implying an additional source of heating around these quasars which is
not yet present in the general IGM at slightly lower redshift.  This
heating is most likely due to the recent reionisation of \HeII in
vicinity of these quasars, which have hard, non-thermal ionising
spectra.  The elevated temperatures may therefore represent evidence
for the earliest stages of \HeII reionisation in the most biased
regions of the high-redshift Universe.  The temperature as a function
of distance from the quasars is consistent with being constant, $\log
(T_{0}/\rm K) \simeq 4.2$, with no evidence for a line-of-sight
thermal proximity effect.  However, the limited extent of the quasar
near-zones prevents the detection of \HeIII regions larger than $\sim
5$ proper Mpc.  Under the assumption the quasars have reionised the
\HeII in their vicinity, we infer the data are consistent with an
average optically bright phase of duration in excess of
$10^{6.5}\rm\,yr$.  These measurements represent the highest-redshift
IGM temperature constraints to date, and thus provide a valuable data
set for confronting models of \HI reionisation.

\end{abstract}
 

%%%%%%%%%%%%%%%%%%%%%%%%%%%%%%%%%%%%%%%%%%%%%%%%%%%%%%%%%%%%%%%%%%%%%	
%%%%%%%%%%%%%%%%%%%%%%%%%% SECTION 1 %%%%%%%%%%%%%%%%%%%%%%%%%%%%%%%%
%%%%%%%%%%%%%%%%%%%%%%%%%%%%%%%%%%%%%%%%%%%%%%%%%%%%%%%%%%%%%%%%%%%%%

\section{Introduction}

The epoch of reionisation marks the time when luminous sources became
a dominant influence on the physical state of the intergalactic medium
(IGM).  Studying the impact of reionisation on the IGM therefore
provides valuable insight into the formation and evolution of the
first stars and galaxies (see \citealt{MoralesWyithe10} for a recent
review).  One quantity which is closely related to the timing and
extent of the reionisation epoch is the temperature of the low-density
IGM. The IGM temperature is set primarily by photo-heating and
adiabatic cooling at $z\la 10$, and therefore depends on the spectral
shape of the radiation from early ionising sources and the time
elapsed since intergalactic hydrogen and helium were reionised ({\it
  e.g.}  \citealt{MiraldaRees94,Theuns02,HuiHaiman03}).

Existing measurements of the IGM temperature are obtained from the
forest of intergalactic \HI \Lya absorption lines observed in the
spectra of bright quasars.  The widths of the \Lya absorption lines
are sensitive to the temperature of the IGM through a combination of
thermal (Doppler) broadening and pressure (Jeans) smoothing of the
underlying gas distribution ({\it e.g.}
\citealt{Haehnelt98,Peeples10a}), in addition to broadening from
peculiar motions and the Hubble flow ({\it e.g.}
\citealt{Hernquist96,Theuns00}).  Consequently, using a statistic
sensitive to the thermal broadening kernel combined with an accurate
model for measurement calibration (typically high resolution
hydrodynamical simulations of the IGM), various authors have placed
constraints on the thermal evolution of the IGM at $2\leq z \leq 4.8$
using the \Lya forest
(\citealt{Schaye00,Ricotti00,TheunsZaroubi00,McDonald01,Zaldarriaga02,Lidz10,Becker11}).

\cite{Becker11} recently presented the most precise IGM temperature
measurements to date obtained from the \Lya forest using a set of 61
high-resolution quasar spectra over a wide redshift range, $2\leq z
\leq 4.8$.  These authors found that the IGM temperature at mean density,
$T_{0}$, is $\sim$8000\,K at $z \simeq
4.4$ and gradually increases toward lower redshift.   \cite{Becker11} attributed this heating
to the onset of an extended epoch of \HeII reionisation driven by
quasars, which produce the hard ionising photons necessary for doubly
ionising intergalactic helium
(\citealt{Madau99,FurlanettoOh08b,McQuinn09}).  When combined with
measurements of the intergalactic \HeII \Lya opacity at $z\la 3$
(e.g. \citealt{Fechner06,Shull10,Worseck11,Syphers11}), these data suggest \HeII
reionisation was in its final stages by or after $z\simeq 3$.

However, in order to probe deeper into the epoch of \HI reionisation
at $z\geq 6$, constraints on the IGM temperature at higher redshift
are required.  Unfortunately, due to the increasing \Lya opacity of
the IGM (\citealt{Songaila04,Fan06,Becker07}), measurements of the IGM
temperature from the general \Lya forest are extremely challenging at
$z>5$.  \cite{Bolton10} (hereafter B10) recently side-stepped this
difficulty by measuring the IGM temperature in the vicinity of the
$z=6.02$ quasar SDSS J0818$+$1722.  Using detailed numerical
simulations, B10 demonstrated that the cumulative probability
distribution function (CPDF) of Doppler parameters measured from \Lya
absorption lines in the highly ionised near-zone is sensitive to
in-situ photo-heating by the quasar, as well as heating from earlier
ionising sources.  B10 therefore used the Doppler parameter CPDF to
infer the IGM temperature at mean density within $\sim5$ proper Mpc of
the quasar, $\log (T_{0}/\rm\,K)=4.37^{+0.09}_{-0.15}$, at 68 per cent
confidence.  However, the analysis of only a single line-of-sight
means the constraint presented by B10 has a rather large statistical
uncertainty, and is not a reliable representation of the IGM thermal
state at $z\simeq 6$ as a whole.  Temperature measurements from
independent lines-of-sight would therefore significantly aid in
reducing the statistical uncertainty on an averaged measurement, as
well as enabling an investigation of line-of-sight temperature
variations.

The goal of this study is to extend and improve upon the preliminary
work of B10 in several important ways.  Firstly, we analyse the line
widths in the near-zones of six further quasars at $z \simeq 6$ using
high-resolution data obtained with Keck/HIRES and Magellan/MIKE
(\citealt{Becker06,Becker07,Calverley11}).  These are combined with
improved measurements of the quasar systemic redshifts and absolute
magnitudes (\citealt{Carilli10}).  Secondly, we significantly expand
the suite of numerical simulations used to calibrate our temperature
measurements, enabling us to explore a wider range of thermal
histories.  Finally, we undertake a more detailed analysis of the key
systematic uncertainties identified in B10: metal line contamination
and continuum placement.  In addition, we also now investigate the
impact of the uncertain thermal history of the IGM at $z>6$ on the
temperature measurements.

This paper is structured as follows.  In Section 2 we present the
observational data and numerical simulations used in this work.  In
Section 3, we briefly review the method for obtaining the near-zone
temperature constraints, which was originally discussed in detail by
B10.  We investigate the important systematic uncertainties on our
measurements in Section 4, and in Section 5 we present our temperature
measurements before finally concluding in Section 6.  An appendix
which presents some tests of our temperature measurement procedure is
given at the end of the paper.  Throughout we assume $\Omega_{\rm
  m}=0.26$, $\Omega_{\Lambda}=0.74$, $\Omega_{\rm b}=0.0444$,
$h=0.72$, $\sigma_{8}=0.80$, $n_{\rm s}=0.96$ (\citealt{Komatsu11})
and a helium fraction by mass of $Y=0.24$ (\citealt{OliveSkillman04}).
All distances are quoted in proper units unless otherwise stated.

%%%%%%%%%%%%%%%%%%%%%%%%%%%%%%%%%%%%%%%%%%%%%%%%%%%%%%%%%%%%%%%%%%%%%	
%%%%%%%%%%%%%%%%%%%%%%%%%% SECTION 2 %%%%%%%%%%%%%%%%%%%%%%%%%%%%%%%%
%%%%%%%%%%%%%%%%%%%%%%%%%%%%%%%%%%%%%%%%%%%%%%%%%%%%%%%%%%%%%%%%%%%%%

\section{Data and numerical modelling}
\subsection{Observational data}

\begin{table*}
  \centering
  \caption{The seven quasar spectra analysed in this study.  The
    columns list, from left to right, the quasar name, redshift and
    absolute AB magnitude at $1450\rm\,\AA$, the extent of the region
    (in $\rm km\,s^{-1}$ and proper Mpc) around the quasar where \Lya
    Voigt profiles are fitted to the spectrum, and the mean
    transmission and average signal-to-noise within this region.
    Further details of the observations and references are provided in
    the remaining columns.}
    \begin{tabular}{c|c|c|c|c|c|c|c|c|c|c|c}
 
      \hline
   Name   & $z_{\rm  q}$$^{\rm a}$  & $M_{1450}$$^{\rm a}$  &   $v_{\rm H,fit}$ & $R_{\rm fit}$ &$\langle F \rangle$ & S/N & Inst. & Dates & $t_{\rm exp}$  & Refs.\\
   & & &  [$\rm km\,s^{-1}$] & [Mpc] & & & & & [hrs] & \\
   \hline
    SDSS J1148+5251 & 6.4189 & -27.82 & 250--3250 & 0.3--4.4 & 0.471  & 23 & HIRES & Jan 2005--Feb 2005 & 14.2 & 1 \\
    SDSS J1030+0524 & 6.308  & -27.16 & 250--3450 & 0.3--4.8 & 0.499  & 24 & HIRES & Feb 2005           & 10.0 & 1 \\
    SDSS J1623+3112 & 6.247  & -26.67 & 250--3150 & 0.3--4.4 & 0.573  & 21 & HIRES & Jun 2005           & 12.5 & 1 \\
    SDSS J0818+1722 & 6.02   & -27.40 & 250--4105 & 0.4--6.0 & 0.569  & 13 & HIRES & Feb 2006           & 8.3  & 2 \\
    SDSS J1306+0356 & 6.016  & -27.19 & 250--3400 & 0.4--5.0 & 0.490  & 21 & MIKE  & Feb 2007           & 6.7  & 3 \\
    SDSS J0002+2550 & 5.82   & -27.67 & 250--3150 & 0.4--4.8 & 0.586  & 41 & HIRES & Jan 2005--Jul 2008 & 14.2 & 1,3 \\
    SDSS J0836+0054 & 5.810  & -27.88 & 250--3100 & 0.4--4.7 & 0.556  & 27 & HIRES & Jan 2005           & 12.5 & 1 \\
    
    \hline
    \multicolumn{10}{|l|}{$^{\rm a}$ Redshift and absolute magnitudes are from \citet{Carilli10}.} \\
    \multicolumn{10}{|l|}{References:} \\
   \multicolumn{10}{|l|}{1 - \citet{Becker06}; 2 - \citet{Becker07}; 3 - \citet{Calverley11}} \\
\end{tabular}
\label{tab:obsdata}
\end{table*}

\begin{figure*}
\centering
\begin{minipage}{180mm}
\begin{center}
\psfig{figure=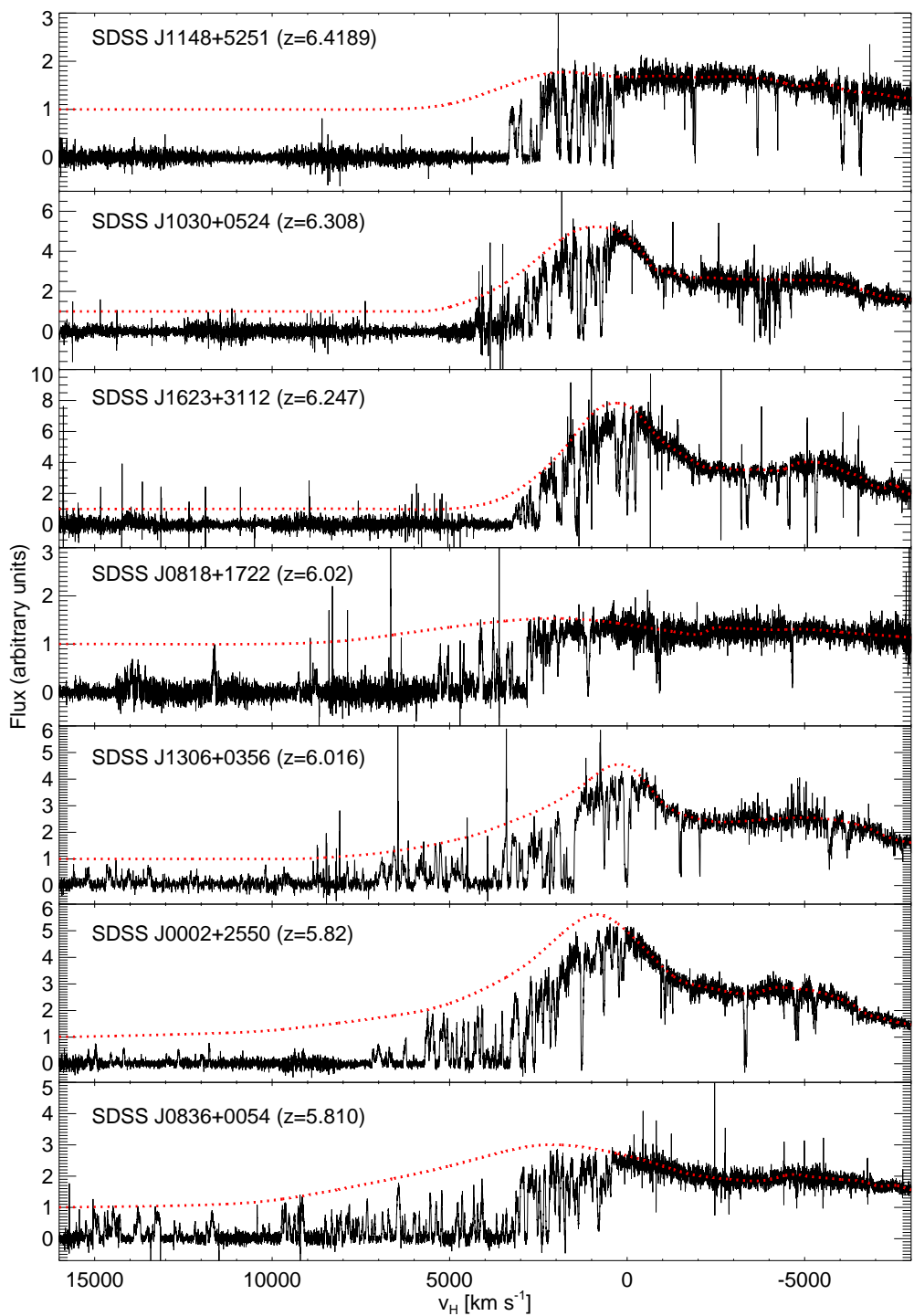,width=0.8\textwidth}
\vspace{-0.95cm}
\caption{The seven quasar spectra analysed in this work against
  velocity with respect to the rest frame \Lya wavelength.  The
  spectra have already been normalised by a power-law with $f_{\rm
    \nu} \propto \nu^{-0.5}$ (see text for details).  A slowly varying
  spline (dotted red curves) has also been fitted to the \Lya emission
  lines in each spectrum to approximate the quasar continuum.  A
  detailed analysis of the systematic uncertainty associated with the
  continuum fitting procedure is presented in Section 4.2.}
\label{fig:contfits}
\end{center}
\end{minipage}
\end{figure*}

\begin{figure*}
\centering
\begin{minipage}{180mm}
\begin{center}
\psfig{figure=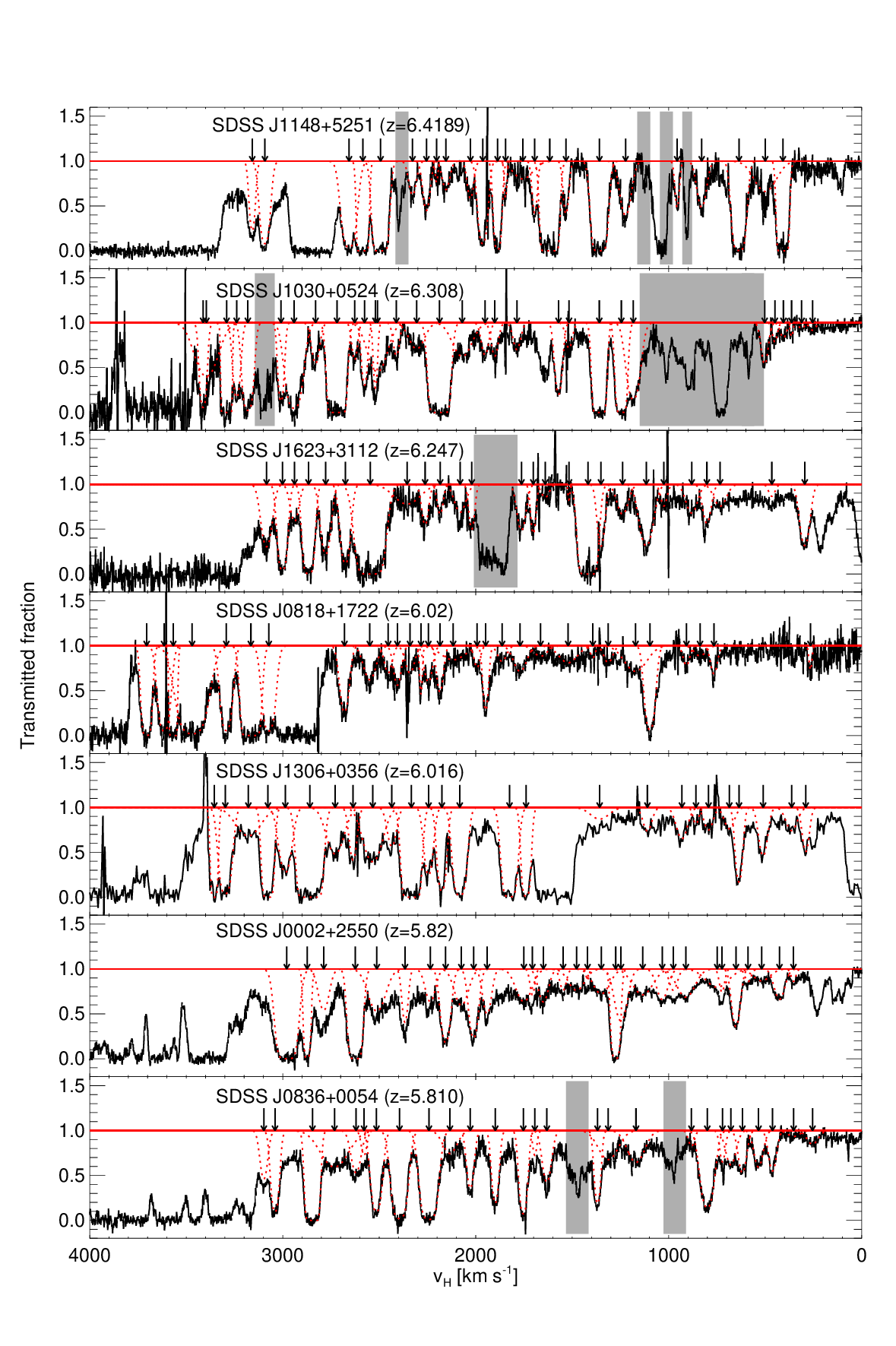,width=0.8\textwidth}
\vspace{-0.95cm}
\caption{Transmitted fraction against the Hubble velocity blueward of
  the \Lya emission line for the seven quasar near-zones analysed in
  this work.  The red dotted curves display the Voigt profile fits to
  the absorption lines made using VPFIT, with the centre of the lines
  marked by the downward pointing arrows.  The grey shaded regions are
  identified as -- or suspected of -- containing metal lines, and are
  thus excluded from the analysis (see Section~\ref{sec:metals} for
  further details).}
\label{fig:linefits}
\end{center}
\end{minipage}
\end{figure*}

We analyse seven high-resolution quasar spectra in this work.  The
spectra for six of these objects were obtained using the High
Resolution Echelle Spectrograph (HIRES, \citealt{Vogt94}) on the 10-m
Keck I telescope using a 0.86\arcsec slit, which produces a resolution
of $R = 40\,000$ (FWHM = $6.7\rm\,km\,s^{-1}$).  The seventh spectrum
was obtained using the Magellan Inamori Kyocera Echelle (MIKE,
\citealt{Bernstein03}) spectrograph on the 6.5-m Magellan II
telescope.  This instrument has a slightly lower resolution compared
to HIRES, with FWHM=$13.6\rm\,km\,s^{-1}$.  The data were processed
using a custom set of IDL routines (\citealt{Becker06,Becker07}) that
include optimal sky subtraction (\citealt{Kelson03}). The final binned
pixel sizes are 2.1~km~s$^{-1}$ for the HIRES spectra, and
$5.0\rm\,km\,s^{-1}$ for the MIKE data. 

The quasars are fully summarised in Table~\ref{tab:obsdata}, and the
spectra are displayed in Figures~\ref{fig:contfits}
and~\ref{fig:linefits} .  The systemic redshifts and absolute AB
magnitudes are taken from \cite{Carilli10}.  The continuum fitting in
the near-zones was performed by first dividing the spectra by a
power-law with $f_{\nu} \propto \nu^{-0.5}$, normalised near
$1280\,(1+z)$~\AA.  The results of this procedure are displayed in
Figure~\ref{fig:contfits}.  The \Lya emission line was then fitted
with a slowly varying spline, shown by the dotted red curves in
Figure~\ref{fig:contfits}.  The fully normalised near-zone spectra are
displayed in Figure~\ref{fig:linefits}.  Further details on the
continuum fitting procedure may be found in B10, and the impact of the
continuum fitting uncertainties on our results is examined in detail
in Section~\ref{sec:cont}.

\begin{figure}
\begin{center}
  \includegraphics[width=0.45\textwidth]{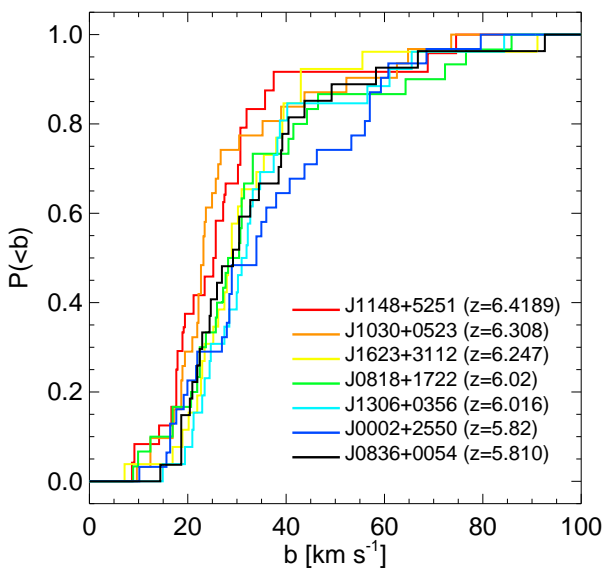}
\vspace{-0.3cm}
 \caption{The Doppler parameter CPDFs obtained from the near-zones of
   the seven quasars analysed in this study.  The Doppler parameters
   are obtained by fitting Voigt profiles to the \Lya absorption in
   the quasar near-zones, and exclude fits in regions suspected of
   containing metal absorption lines.}
\label{fig:bcpdf}
\end{center}
\end{figure}

We obtain the Doppler parameter CPDF for the \Lya absorption lines
following a procedure identical to the approach described by B10.  We
use an automated version of the Voigt profile fitting package {\sc
  VPFIT}\footnote{http://www.ast.cam.ac.uk/$\sim$rfc/vpfit.html}, to
fit lines over the velocity ranges indicated in
Table~\ref{tab:obsdata}.  All lines with $\log (N_{\rm HI}/\rm
cm^{-2})>17$, $b>100\rm\,km\,s^{-1}$ and with relative errors in
excess of 50 per cent are discarded for the temperature
analysis. These discarded lines make up 30 per cent of the total
fitted to the data.  The remaining line fits are indicated by the red
dotted curves in Figure~\ref{fig:linefits}, with the centre of each
line indicated by the downward pointing arrows.  The resulting Doppler
parameter CPDFs for each of the seven quasars are displayed in
Figure~\ref{fig:bcpdf}.

\subsection{Hydrodynamical simulations} \label{sec:sims}

The synthetic \Lya absorption spectra used to calibrate the Doppler
parameter CPDF measurements are constructed using high-resolution
hydrodynamical simulations combined with a line-of-sight Lyman
continuum radiative transfer algorithm.  The hydrodynamical
simulations were performed using the parallel Tree-SPH code {\sc
  GADGET-3}, which is an updated version of the publicly available
code {\sc GADGET-2} (\citealt{Springel05}).  The simulations have a
box size of $10h^{-1}$ comoving Mpc and a gas particle mass of
$9.2\times 10^{4}h^{-1}M_{\odot}$.  Outputs are obtained from the
simulations at $z=6.42$, $z=6.28$, $z=6.01$ and $z=5.82$.  Our
procedure for constructing simulated quasar spectra along with the
appropriate numerical convergence tests are discussed in detail by
B10.  For brevity, we focus instead on describing the additional
improvements made for this study.

In order to explore the effect of a wide range of gas temperatures on
the Doppler parameter CPDF, we perform an expanded set of
hydrodynamical simulations which employ a variety of different thermal
histories.  We use 14 hydrodynamical simulations in total, which are
summarised in Table~\ref{tab:sims}.  The initial gas temperature in
these simulations, prior to any quasar heating, is set using a
pre-computed ultra-violet (UV) background model.  The fiducial UV
background in this study is the galaxies and quasars emission model of
\cite{HaardtMadau01}, in which the IGM is reionised instantaneously at
$z=9$.  In models A through to M, we construct simulations with
different initial temperatures by rescaling the \cite{HaardtMadau01}
photo-heating rates with a constant factor, $\zeta$, such that
$\epsilon_{\rm i} = \zeta \epsilon_{\rm i}^{\rm HM01}$.  Here
$\epsilon_{\rm i}^{\rm HM01}$ are the \cite{HaardtMadau01}
photo-heating rates for species $i=[\rm H\,\scriptstyle I$, $\rm
  He\,\scriptstyle I$, $\rm He\,\scriptstyle II$].

\begin{figure}
\begin{center}
  \includegraphics[width=0.45\textwidth]{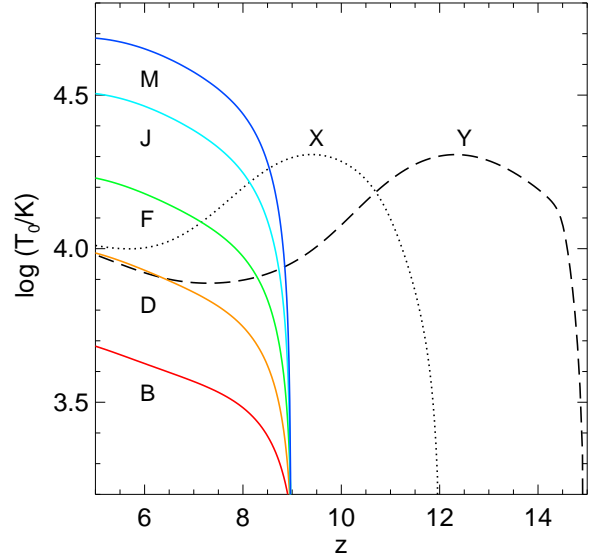}
\vspace{-0.3cm}
\caption{The temperature at mean density and its dependence on
  redshift for a selection of the hydrodynamical simulations listed in
  Table~\ref{tab:sims}.  The solid curves display the simulations
  using our fiducial thermal history which is based on the galaxies
  and quasars UV background of HM01.  We also explore the effect of a
  more extended period of heating on our results with two additional
  models, X and Y.  These are shown by the dotted and dashed curves,
  where the IGM is first heated at $z=12$ and $z=15$, respectively.}
\label{fig:hydrosims}
\end{center}
\end{figure}

The temperature at mean density as a function of redshift in a
selection of these hydrodynamical simulations, prior to any quasar
heating, is displayed in Figure~\ref{fig:hydrosims}.  The
photo-heating is coupled to the hydrodynamical response of the gas in
the simulations; different heating histories will therefore result in
a different pressure smoothing scale for each model
(e.g. \citealt{GnedinHui98,Pawlik09}).  Since the \Lya line widths are
sensitive to the changes in the gas temperature through Jeans
smoothing {\it in addition} to thermal broadening
(\citealt{Peeples10a,Becker11}), we also consider two further
customised UV background models, X and Y, for which photo-heating
begins at $z=12$ and $z=15$, respectively.  These models are displayed
as the dotted and dashed curves in Figure~\ref{fig:hydrosims}.  These
extended heating histories will result in gas being pressure smoothed
on larger scales, even if the instantaneous gas temperature is similar
to our fiducial models.  The models are chosen to provide a plausible
upper limit to the Jeans smoothing scale at $z\simeq 6$, and we shall
use them to explore the effect of uncertainties in the IGM thermal
history on our results in Section~\ref{sec:Jeans}.

\subsection{Lyman continuum radiative transfer}

We next include the impact of photo-heating by the quasar on the
surrounding IGM using our line-of-sight radiative transfer algorithm.
For each of the seven observed quasars, we first generate a set of
synthetic lines-of-sight using the simulation output closest to the
quasar systemic redshift.  A total of 100 lines-of-sight of length
$55h^{-1}$ comoving Mpc are extracted around the most massive haloes
identified from {\it each} of the hydrodynamical simulations and for
{\it each quasar}, yielding skewers through the IGM density,
temperature and peculiar velocity field.  For our fiducial thermal
history, this yields a total of $1200$ simulated lines-of-sight for
calibrating each of the seven observed Doppler parameter CPDFs.

The next stage is to compute the transfer of ionising radiation along
the lines-of-sight.  We assume the quasar spectra are described by a
broken power law, $f_{\nu}\propto \nu^{-0.5}$ for $1050{\rm\,\AA}
<\lambda < 1450\,\rm \AA$ and $f_{\nu}\propto \nu^{-\alpha_{\rm q}}$
for $\lambda<1050\,\rm \AA$.  We adopt a fiducial extreme-UV (EUV)
spectral index of $\alpha_{\rm q}=1.5$ in this work, consistent with
radio quiet quasars at lower redshift (e.g. \citealt{Telfer02}).  The
optically bright phase of the quasars is assumed to be $t_{\rm
  q}=10^{7}\rm\,yr$ (\citealt{Haehnelt98b,Croton09}).  We assume the
\HI and \HeI around the $z\simeq 6$ quasars is already highly ionised
when the quasar turns on ({\it e.g.}  \citealt{Wyithe08}, B10), but
that helium has yet to be doubly ionised. The subsequent reionisation
and photo-heating of \HeII by the quasar thus results in an additional
IGM temperature increase of $7000$--$9000\rm\,K$.  Note, however, an
EUV spectral index which is softer (harder) than $\alpha_{\rm q}=1.5$
will decrease (increase) the amount of \HeII photo-heating
(\citealt{Bolton09,McQuinn09}). The IGM temperatures at mean density
prior to and after \HeII heating by the quasar are summarised in
Table~\ref{tab:sims}.

\begin{table}
  \centering
  \caption{The hydrodynamical simulations used in this work.  From
    left to right, the columns list the simulation name, the scaling
    factor for the UVB photo-heating rates (see main text for
    details), the redshift which the gas is first photo-heated in the
    hydrodyamical simulations and the volume weighted gas temperature
    at mean density, both prior to and after \HeII photo-heating by
    the quasar.  The temperatures in this instance are shown for the
    simulations of SDSS J0818$+$1722 at $z=6.02$, but are broadly
    similar for the other six quasars. }

  \begin{tabular}{c|c|c|c|c}
   
    \hline
    Model  & $\zeta$ & $z_{\rm heat}$ & $\log (T_{\rm 0}^{\rm initial}$/K) & $\log (T_{\rm 0}$/K) \\  
    \hline
    A  & 0.1 & 9.0  & 3.29  & 4.03 \\
    B  & 0.3 & 9.0  & 3.62  & 4.11 \\
    C  & 0.5 & 9.0  & 3.78  & 4.16 \\  
    D  & 0.8 & 9.0  & 3.93  & 4.22 \\
    E  & 1.1 & 9.0  & 4.02  & 4.27 \\
    F  & 1.8 & 9.0  & 4.17  & 4.35 \\
    G  & 2.6 & 9.0  & 4.28  & 4.42 \\
    H  & 3.6 & 9.0  & 4.37  & 4.49 \\
    J  & 4.7 & 9.0  & 4.45  & 4.54 \\
    K  & 5.9 & 9.0  & 4.51  & 4.59 \\
    L  & 7.8 & 9.0  & 4.59  & 4.65 \\
    M  & 9.3 & 9.0  & 4.63  & 4.69 \\
    X  & Varied & 12.0 & 4.00  & 4.26 \\
    Y  & Varied & 15.0 & 3.92  & 4.22 \\
    \hline
  \end{tabular}
\label{tab:sims}
\end{table}

%%%%%%%%%%%%%%%%%%%%%%%%%%%%%%%%%%%%%%%%%%%%%%%%%%%%%%%%%%%%%%%%%%%%%	
%%%%%%%%%%%%%%%%%%%%%%%%%% SECTION 3 %%%%%%%%%%%%%%%%%%%%%%%%%%%%%%%%
%%%%%%%%%%%%%%%%%%%%%%%%%%%%%%%%%%%%%%%%%%%%%%%%%%%%%%%%%%%%%%%%%%%%%

\section{Methodology} \label{sec:method}

We now briefly review our method for obtaining the near-zone
temperature constraints from the Doppler parameter CPDF (but see B10
for further details).  The advantage of using the Doppler parameter
CPDF is that it fully uses the limited number of absorption lines
available in the observational data and avoids the loss of information
associated with binning.  Although not all the absorption lines in the
CPDF are purely thermally broadened (\citealt{Theuns00}), the thermal
broadening kernel nevertheless acts on all the \Lya absorption and the
entire CPDF therefore remains sensitive to changes in the gas
temperature.  The second advantage is that we may compare the observed
CPDFs to our simulations using the ``D-statistic'', which is very
similar to the parameter used in a Kolmogorov-Smirnov test ({\it e.g.}
\citealt{Press92}).  This approach has the advantage of providing a
non-parametric measure which quantifies the difference between Doppler
parameter CPDFs drawn from different models and the data.  On the
other hand, obtaining absolute confidence intervals on the temperature
using this method requires a large set of realistic simulations for
calibrating the D-statistic, and these simulations are time consuming
to perform and analyse.

For each quasar, we therefore use our sets of synthetic spectra to
construct Monte Carlo realisations of the Doppler parameter CPDF for
each of the fiducial simulations, A through to M.  The simulated CPDFs
are obtained by performing exactly the same line fitting procedure
used on the observational data.  For each simulated line-of-sight in
the set of 100 spectra, we then measure the D-statistic, which is the
maximum difference between the Doppler parameter CPDF for that
simulated line-of-sight and the Doppler parameter CPDF for all 100
simulated lines-of-sight:

\begin{equation} D_{\rm i}= {\rm max} | P(<b)_{\rm i} - P(<b)_{\rm all}|, \,i=1...100,\end{equation}

\noindent
where we preserve the sign of the difference.  The D-statistic CPDF
for a model with known temperature $\log T_{0}$, $P(<D|\log T_{0})$,
can then be constructed.  We then smooth over the noise arising from
the discrete sampling of the D-statistic CPDF with a Gaussian filter
of width $\sigma=0.025$ before computing its derivative,
$\frac{dP(<D_{\rm obs}|\log T_{0})}{dD}$.  Note that in this work we
fit 100 spectra for each model, as opposed to only 30 in B10, enabling
a finer sampling of the D-statistic distribution.

By then measuring the D-statistic for the observed line-of-sight,
$D_{\rm obs}$, we may then use the D-statistic CPDF along with Bayes'
theorem to obtain a probability distribution for the logarithm of the
{\it observed} temperature at mean density, $\log T_{0}$:

\begin{equation} p(\log T_{0}|D_{\rm obs}) = K \frac{dP(<D_{\rm obs}|\log T_{0})}{dD} p(\log T_{0}),  \label{eq:bayes} \end{equation}

\noindent
where $p(\log T_{0})$ is the prior on $\log T_{0}$ and $K$ is a
constant which normalises the total probability to unity.  As in B10,
we adopt a flat prior, $p(\log T_{0})$, but adopt a more extended
prior temperature range, $3.2 \leq \log (T_{0}/\rm K) \leq 4.8$ using
our expanded set of simulations.  The prior is chosen to encompass the
range of full range of initial and final temperatures in our
simulations (see Table 2) and is intended to represent a reasonable
range for the IGM temperature following \HeII and/or \HI reionisation
(see {\it e.g.} \citealt{Trac08,McQuinn09}).  We then evaluate
$\frac{dP(<D_{\rm obs}|\log T_{0})}{dD}$ at twelve discrete points
using each of our fiducial hydrodynamical simulations, and use a cubic
spline to interpolate between these to obtain a continuous
distribution.  Lastly, we infer confidence intervals around the median
$\log T_{0}$ by integrating $p(\log T_{0}|D_{\rm obs})$ over the
appropriate limits.  All our temperature measurements are therefore
quoted as the {\it median} of the $p(\log T_{0}|D_{\rm obs})$
distribution with the confidence intervals around the median.  This
will be implicit throughout the remainder of the paper.

%%%%%%%%%%%%%%%%%%%%%%%%%%%%%%%%%%%%%%%%%%%%%%%%%%%%%%%%%%%%%%%%%%%%%
%%%%%%%%%%%%%%%%%%%%%%%%%% SECTION 4 %%%%%%%%%%%%%%%%%%%%%%%%%%%%%%%%
%%%%%%%%%%%%%%%%%%%%%%%%%%%%%%%%%%%%%%%%%%%%%%%%%%%%%%%%%%%%%%%%%%%%%

\section{Systematic uncertainties}

We now turn to examine the key systematic uncertainties in our
analysis.  In B10, we considered five potential sources of systematic
error.  The first three -- the mean transmitted fraction in the quasar
near-zone, the influence of strong galactic winds, and the dependence
of the IGM temperature on gas density, were found to have a negligible
effect on the Doppler parameter CPDF.  In the latter case, this is
because transmission in the $z\simeq 6$ near-zones is sensitive to gas
close to mean density.  On the other hand, B10 estimated that metal
line contamination and uncertainties in the continuum placement could
systematically bias temperature measurements by up to $\sim
2000\rm\,K$.  Although these uncertainties are small compared to the
statistical uncertainty on the B10 measurement for a single
line-of-sight, they will be important to consider for the larger data
set used in this study.  We furthermore examine an additional and
important systematic which was not included in B10 -- the effect of
the uncertain thermal history at $z>6$ ({\it e.g.}
\citealt{Becker11}).

\subsection{Metal lines} \label{sec:metals}

We first examine the effect of metal line contamination on our
results.  As our synthetic spectra do not include absorption from
intervening metals, we must remove any metal contamination present in
the observational data in order to avoid biasing our temperature
measurements.  The erroneous identification of metal lines as \HI \Lya
absorption can lead to underestimated temperatures, since in general
metal ions exhibit significantly narrower line widths compared to the
\Lya lines.

Our approach is to identify possible metal lines in the quasar
near-zones and exclude these regions from our Doppler parameter
analysis.  Potential contaminants lying within
the near-zones were first identified  by searching for metal line systems at other
wavelengths in the quasar spectra, and flagging any associated lines
within the near-zones.  This was followed by flagging any absorption
features in the near-zones that subsequently remained unidentified and
appeared too narrow to be \HI \Lya lines.  This procedure is most
reliable for quasars which have near-IR spectra, enabling greater
coverage redward of the \Lya emission line.  Near-IR data was
available for five of the seven quasars analysed in this work; the
exceptions are J1623$+$3112 and J1306$+$0356.

\begin{figure*}
\centering
\begin{minipage}{180mm}
\begin{center}
  \psfig{figure=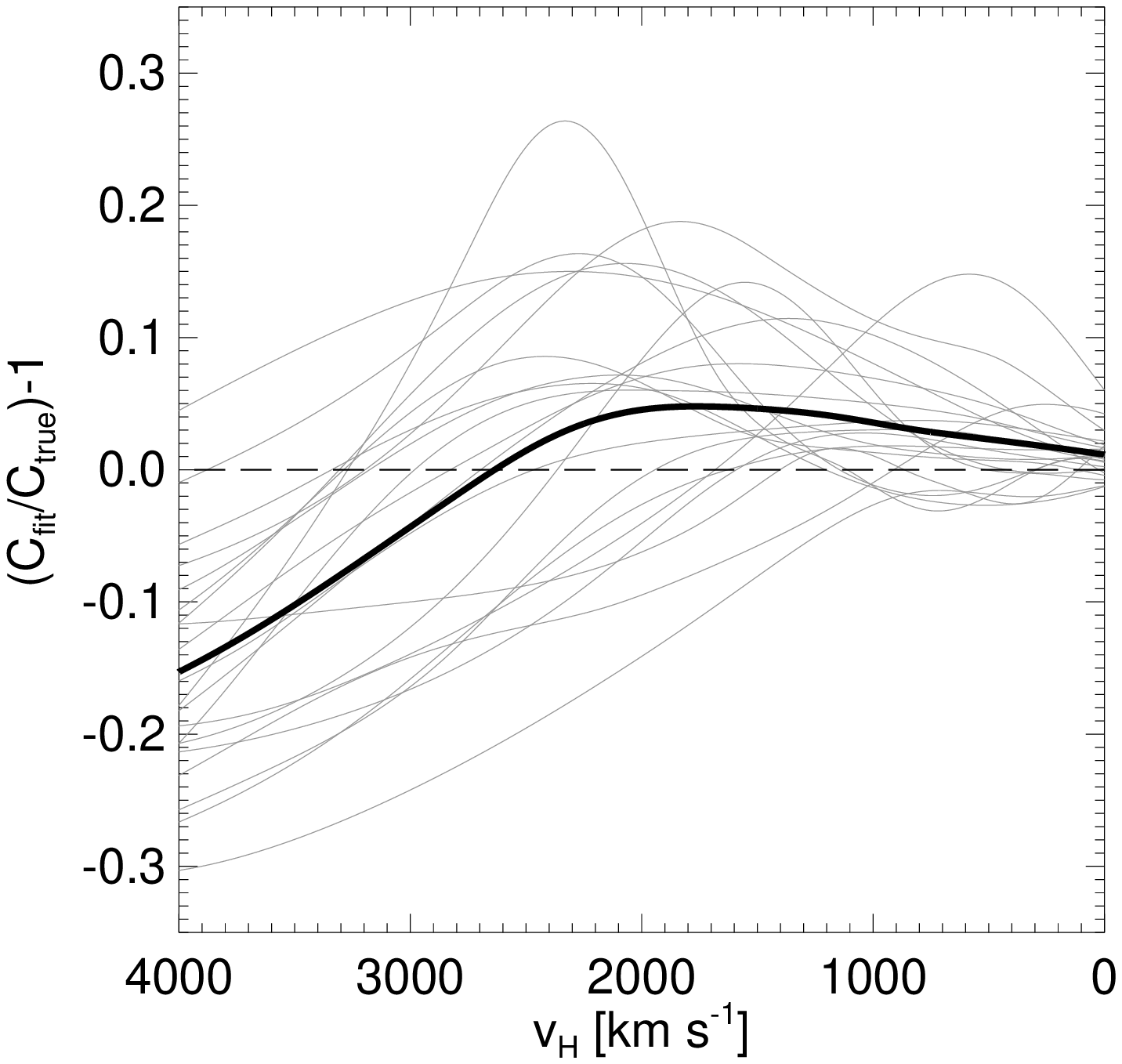,width=0.45\textwidth}
  \psfig{figure=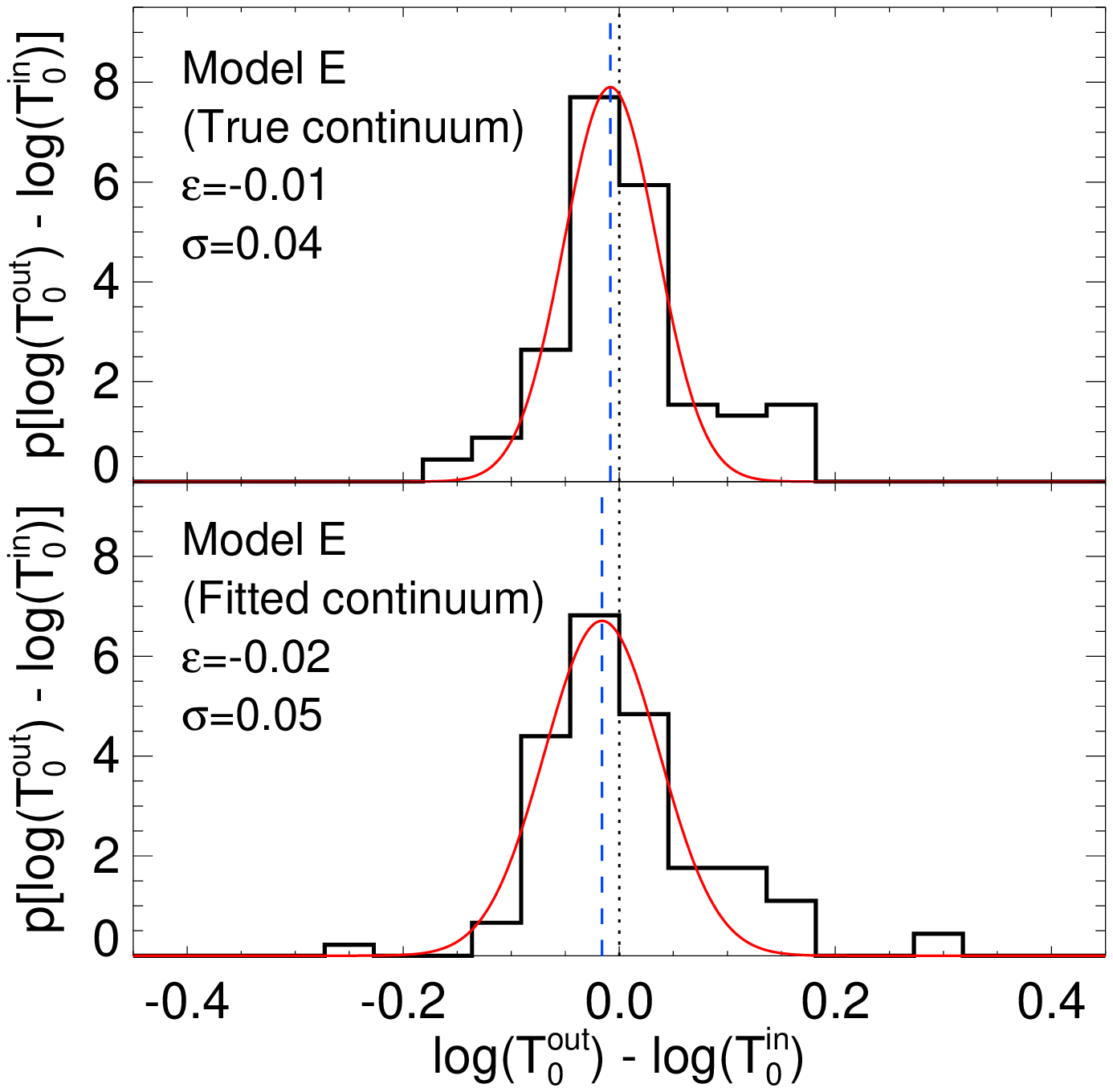,width=0.45\textwidth}
  \vspace{-0.3cm}
  \caption{{\it Left:} The continuum level, $C_{\rm fit}$, recovered
    from a blind analysis of $20$ synthetic spectra relative to the
    true unabsorbed continuum, $C_{\rm true}$, against Hubble velocity
    blueward of rest-frame Ly$\alpha$.  The light grey curves display
    the relative continuum uncertainty for each spectrum, while the
    thick black line shows the average. {\it Upper right:} The
    distribution of the difference between the temperature recovered
    using the algorithm detailed in Section~\ref{sec:method} and the
    input value, $\log T_{0}^{\rm out}- \log T_{0}^{\rm in}$ (black
    solid lines), for 100 synthetic spectra drawn from model E where
    the true continuum is known exactly.  The resulting distribution
    may be approximated by a Gaussian with mean $\epsilon = -0.01$ dex
    and standard deviation $\sigma=0.04$ dex (red solid curve),
    indicating that the true temperature is recovered well with some
    intrinsic scatter. The vertical black dotted line and blue dashed
    lines mark the zero offset and mean of the best fit Gaussian,
    respectively.  {\it Lower right:} The distribution of $\log
    T_{0}^{\rm out}- \log T_{0}^{\rm in}$ for the same spectra after
    randomly dividing each by one of the error functions in the
    left-hand panel.  The distribution may be approximated as a
    Gaussian with an offset of $\epsilon=-0.02$ dex and standard
    deviation $\sigma=0.05$ dex. }
\label{fig:continuum}
\end{center}
\end{minipage}
\end{figure*}

The contaminated regions identified in the near-zones are marked by
the grey shaded regions in Figure~\ref{fig:linefits}, and are excluded
from our Doppler parameter CPDF analysis.  For three of the quasars,
J0818$+$1722, J1306$+$0356 and J0002$+$2550, no metal lines were
identified.  Removing the metal contamination slightly raises the
temperature constraints obtained from the other four spectra,
producing an increase of $\sim 0.02$ dex in our $\log T_{0}$
constraints.  Note, however, that it is possible that metal lines
which are highly blended with the \Lya lines remain unidentified.
Throughout the rest of the paper, we therefore conservatively add an
additional scatter of $0.02$ dex in quadrature to our final $\log
T_{0}$ constraints for each line-of-sight.

\subsection{Continuum placement} \label{sec:cont}

The second important systematic identified in B10 was the effect of
continuum placement uncertainties on the temperature measurements.  If
the continuum is placed too low (high) on the observational data, the
IGM temperature around the quasars will be underestimated
(overestimated) as the absorption line widths are effectively narrowed
(broadened).  B10 attempted to account for continuum placement
uncertainties by normalising the synthetic spectra by the highest
transmitted flux in segments of length $1000\rm\,km\,s^{-1}$.  The
motivation for this was to minimise bias by treating the synthetic
data in a similar fashion to the observational data.

However, the drawback of this approach is that it only crudely mimics
the continuum fitting process on the observational data.  In this
study, we instead perform ``blind'' normalisations on the synthetic
spectra to estimate the continuum placement uncertainty (see also
e.g. \citealt{Tytler04,Faucher08}).  Our procedure is as follows.  One
of us (JSB) constructed $20$ different simulated spectra with
sufficient wavelength coverage on either side of the \Lya emission
line for normalisation.  Each spectrum was then multiplied with one of
the continuum fits compiled by \cite{KramerHaiman09}, obtained from a
large sample of low redshift, unobscured quasars.  The spectra were
then processed to resemble the resolution and noise properties of the
quasars analysed in this work.  A second author (GDB), who was
responsible for continuum fitting the observational data, then
proceeded to normalise the synthetic spectra without any prior
knowledge of the true continuum.

The results of the blind continuum fits to the synthetic data are
displayed in the left panel of Figure ~\ref{fig:continuum}, where the
continuum level, $C_{\rm fit}$, recovered from the analysis is shown
relative to the true continuum, $C_{\rm true}$, against Hubble
velocity blueward of rest frame Ly$\alpha$.  The light grey curves
display the relative continuum uncertainty for each of the 20
synthetic spectra, while the thick black line shows the average.  On
average the continuum is placed to within 5 per cent of the true value
within $\sim 3000\rm\,km\,s^{-1}$ of the quasar redshift, where the
uncertain shape of the quasar Ly-$\alpha$ emission line is the largest
uncertainty.  However, the continuum placement is almost always biased
low beyond $>3000\rm\,km\,s^{-1}$ -- by as much as 15 per cent on
average -- where the transmitted flux rarely recovers to the
unabsorbed level.  Fortunately, since the majority of the absorption
lines fitted to the observational data in this work lie within
$3000\rm\,km\,s^{-1}$ of the quasar systemic redshift, this suggests
that the impact of the continuum placement on our temperature
measurements should be relatively modest.

We quantify this in the right-hand panel of
Figure~\ref{fig:continuum}, where the recovery of the IGM temperature
at mean density from the simulated Doppler parameter CPDF is tested.
The upper panel displays the distribution of the difference between
the input temperature, $\log T_{0}^{\rm in}$, and the recovered
temperature, $\log T_{0}^{\rm out}$, for 100 simulated spectra drawn
from model E.  This distribution may be approximated by a Gaussian,
displayed by the red curve.  In this instance the continuum is assumed
to be known exactly, and the input temperature is recovered accurately
(to within $0.01$ dex, with a small scatter of $\sim 0.04$ dex).

For comparison, in the lower right hand panel of
Figure~\ref{fig:continuum} we display the distribution of $\log
T_{0}^{\rm out} - \log T_{0}^{\rm in}$ for the same spectra, but after
they have been divided at random by one of the error functions
displayed in the left hand panel of Figure~\ref{fig:continuum}.  It is
apparent that the temperature recovery is only very mildly impacted by
the continuum placement. The recovered temperatures are biased to
slightly lower temperatures (by $0.02$ dex) with some additional
scatter.  This is a smaller effect than that predicted by B10, who
estimated the continuum fitting process could bias the temperature
measurements by as much as $\sim 2000\rm\,K$.  In order to test this
further, we therefore also performed the same temperature recovery
test on a set of spectra drawn from model E, but this time using the
normalisation procedure used by B10.  This resulted in temperatures
which were biased low by $0.03$ dex.  This suggests that the
approximate method used in B10 slightly over-compensates for the
relatively modest systematic offset in the recovered temperature due
to continuum placement.

We therefore conclude that the uncertain continuum placement will
introduce only a small additional uncertainty to our results.
Nevertheless, we shift all our temperature measurements upward by
$+0.02$ dex to account for the estimated continuum bias, and a further
$0.03$ dex is added in quadrature to our estimated uncertainties for
each line-of-sight.

\subsection{Jeans smoothing and the thermal history} \label{sec:Jeans}

The third and final systematic we examine is the uncertain {\it
  thermal history} of the IGM at $z>6$.  This will impact on the
small-scale structure of the \Lya forest at $z\simeq 6$ through the
effect of Jeans smoothing (e.g. \citealt{GnedinHui98,Pawlik09}).  The
finite amount of time required for gas pressure to respond to changes
in the temperature means that two models with different thermal
histories will be pressure smoothed on different scales, even if the
instantaneous temperature is similar at the quasar redshift.  In
practice, because we rely on the simulations to calibrate the
relationship between the Doppler parameter CPDF and the IGM
temperature, if the true thermal history of the IGM is different to
that assumed in our models, a systematic bias will be imparted to our
measurements.  Unfortunately, since the IGM temperature is
unconstrained at $z>6$, the thermal history introduces an important
uncertainty into our analysis ({\it e.g.}  \citealt{Becker11}).

\begin{figure}
\begin{center}
  \includegraphics[width=0.45\textwidth]{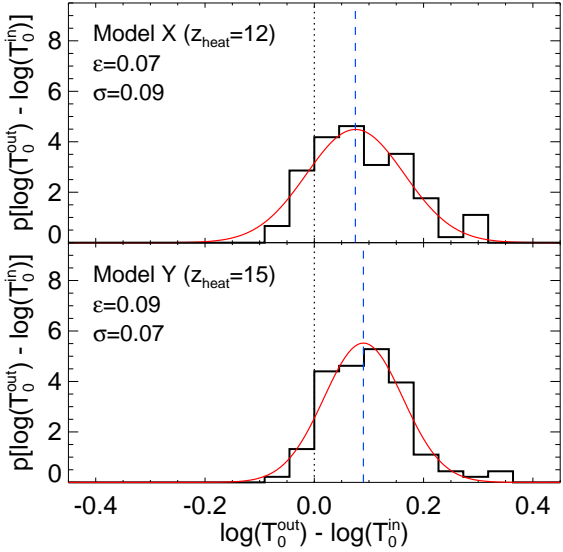}
\vspace{-0.3cm}

\caption{The effect of the uncertain IGM thermal history at $z>6$ on
  the temperatures recovered from quasar near-zones.  {\it Top:} The
  distribution of the offset in the recovered temperature from the
  input value, $\log T_{0}^{\rm out}- \log T_{0}^{\rm in}$ (black
  solid lines), for 100 synthetic spectra drawn from Model X, in which
  photo-heating of the IGM begins at $z=12$.  {\it Bottom:} The
  distribution for model Y, with heating beginning at $z=15$.  In both
  panels, the red curves show the best-fit Gaussian, with mean
  $\epsilon$ and standard deviation $\sigma$ indicated on the panels,
  while the vertical black dotted line and blue dashed lines mark the
  zero offset and the mean of the best-fit Gaussian, respectively.}
  \label{fig:Jeans}
\end{center}
\end{figure}

We investigate the impact of the uncertain thermal history on our
results in Figure~\ref{fig:Jeans}, where we examine the accuracy with
which temperatures are recovered from spectra drawn from models X and
Y.  These models have thermal histories which are more extended than
our fiducial models, with heating beginning at $z=12$ and $z=15$,
respectively (see Figure~\ref{fig:hydrosims}).  The increased amount
of early heating in these two models means that the IGM has had longer to
dynamically respond to the increased temperatures.  As a result, the
gas distribution in these models is physically smoothed on a larger
scale.  In practice, this means that relative to our fiducial
simulations, a greater proportion of the line broadening in spectra
constructed from these simulations will be due the increased physical
extent of the absorbers rather than Doppler broadening.

As a consequence, the recovered distributions for $\log T_{0}^{\rm
  out}- \log T_{0}^{\rm in}$ in Figure~\ref{fig:Jeans} demonstrate
that the inferred temperatures for models X and Y are significantly
higher, by $0.07$--$0.09$ dex, compared to the true temperature in the
models.  The measurements also exhibit more scatter ($\sigma \sim
0.07$--$0.09$ dex) around the mean than observed for the fiducial
models (e.g. the upper right panel in Figure~\ref{fig:continuum}).  As
might be expected, the systematic offset in the recovered temperature
is slightly larger for the model where heating begins at $z=15$. The
very uncertain thermal history $z>6$ therefore means it is possible
that we may overestimate the IGM temperature in the real quasar
near-zones by as much as $0.1$ dex.  On the other hand, if
reionisation occurred very late we may underestimate the temperatures.
However, the latter case will make our conclusions regarding \HeII
photo-heating around the quasars stronger, rather than weaker.  We
therefore address this issue in the next section by presenting
temperature measurements which include an estimate of the impact of
additional Jeans smoothing along with our fiducial results.  These
Jeans smoothing related uncertainties are estimated based on the
results of this analysis: we correct for a systematic offset by
shifting our constraints by $-0.08$ dex and adding an additional
scatter of $\sigma=0.07$ dex in quadrature to the measurement
uncertainties for each line-of-sight.

%%%%%%%%%%%%%%%%%%%%%%%%%%%%%%%%%%%%%%%%%%%%%%%%%%%%%%%%%%%%%%%%%%%%%	
%%%%%%%%%%%%%%%%%%%%%%%%%% SECTION 5 %%%%%%%%%%%%%%%%%%%%%%%%%%%%%%%%
%%%%%%%%%%%%%%%%%%%%%%%%%%%%%%%%%%%%%%%%%%%%%%%%%%%%%%%%%%%%%%%%%%%%%

\section{Results}
\subsection{Line-of-sight temperature constraints}

\begin{figure}
\begin{center}
  \includegraphics[width=0.45\textwidth]{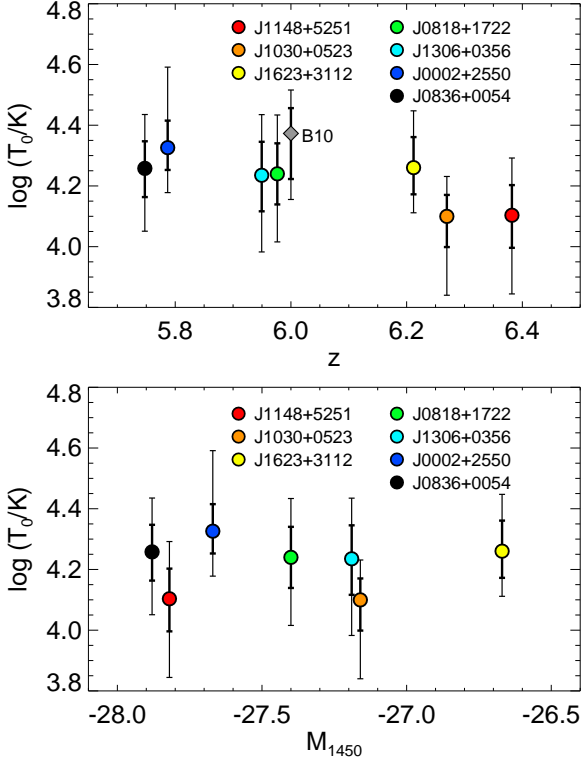}
\vspace{-0.3cm}
\caption{{\it Upper panel:} The filled circles display the recovered
  near-zone temperatures at mean density against redshift for the
  seven quasar spectra analysed in this work.  The thick (thin) error
  bars show the 68 (95) confidence intervals.  The data points for
  J1306$+$0356 and J0836$+$0054 have been offset by $\Delta z=-0.03$
  for clarity.  The grey diamond compares the measurement for
  J0818$+$1722 obtained by B10.  {\it Lower panel:} The near-zone
  temperature at mean density against quasar absolute magnitude at
  $1450\rm\,\AA$.}
\label{fig:temp_single}
\end{center}
\end{figure}

We now present the main results of this study by first considering the
temperature measurements obtained for the individual lines-of-sight.
The temperature measurements at mean density derived for our fiducial
thermal history are summarised in Table~\ref{tab:results}, and are
plotted as a function of redshift (upper panel) and quasar absolute
magnitude (lower panel) in Figure~\ref{fig:temp_single}.  It is
reassuring to note that the temperature constraints mirror the Doppler
parameter CPDFs displayed in Figure~\ref{fig:bcpdf}, with the lowest
temperatures derived from the CPDFs with the smallest Doppler
parameters on average.  The two highest redshift quasars, J1148$+$5251
and J1030$+$0524, exhibit slightly lower temperatures compared to the
rest of the sample, while on the other hand J0002$+$2550 has a
slightly higher temperature.  However, the measurement uncertainties are such
that all seven quasars are consistent with the same IGM temperature
within the 95 per cent confidence intervals.

In the upper panel of Figure~\ref{fig:temp_single}, the temperature
constraints are also compared to the measurement for J0818$+$1722
presented by B10, shown by the grey diamond. Although our revised
measurement for J0818$+$1722 is formally consistent within the 68 per
cent confidence interval, it is systematically lower by $\sim 0.13$
dex.  The first reason for this difference is that in B10 the redshift
of J0818+1722 was given as $z=6.00$ (\citealt{Fan06}).  The revised
redshift of $z=6.02$ presented by \cite{Carilli10} now extends the
region where we fit absorption lines by $855\rm\,km\,s^{-1}$,
resulting in an additional five absorption lines in the near-zone
(yielding a total of 30).  These lines lower the median Doppler
parameter by $\sim 1.7\rm\,km\,s^{-1}$, and shift the temperature
constraint downward by $\sim 0.03$ dex.  The second difference is our
improved treatment of the continuum uncertainty.  As discussed in
Section~\ref{sec:cont}, the approximate continuum uncertainty
correction used in B10 produces temperatures which are $\sim 0.01$ dex
higher compared to our new results.  The third difference is the
extended prior probability used in this work, especially at lower
temperatures.  We may approximate the B10 measurements by using only
models C to J for our analysis and restricting the prior probability
to $4.13\leq \log (T_{0}/\rm K) \leq 4.56$.  This yields a temperature
$\sim 0.03$ dex higher than the value obtained using the full
simulation set.

\subsection{Average temperature constraints}

The primary improvement of this study is the enlarged sample of seven
quasars at $z\simeq 6$, which enables us to significantly improve upon
the large statistical uncertainty in the individual line-of-sight
measurement presented by B10.  The average temperature constraints
derived from the Doppler parameter CPDF for all seven quasars are
therefore given in the final two rows of Table~\ref{tab:results}.  As
discussed in Section~\ref{sec:Jeans}, each column gives the
measurements for both our fiducial thermal history and the case of
additional Jeans smoothing arising from a more extended heating
history.

\begin{figure*}
\centering
\begin{minipage}{180mm}
\begin{center}
  \psfig{figure=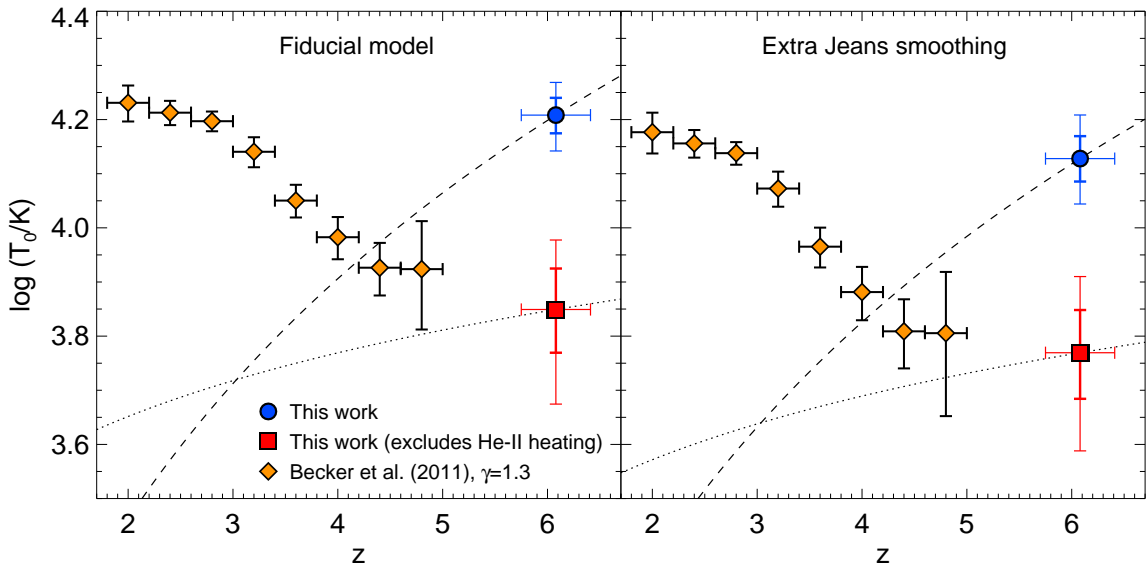,width=0.9\textwidth}
  \vspace{-0.3cm}
  \caption{{\it Left:} The filled blue circle shows the temperature at
    mean density recovered from all seven quasar spectra analysed in
    this work. The thick (thin) error bars display the the 68 (95) per
    cent confidence intervals.  The red square shows the temperature
    constraint after subtracting the expected photo-heating from the
    reionisation of \HeII by the quasars.  For comparison, the open
    diamonds give the temperature measurements from the general \Lya
    forest at $z<5$ obtained by \citet{Becker11} with $2\sigma$
    uncertainties.  The slope for the IGM temperature-density
    relation, $T=T_{0}(1+\delta)^{\gamma-1}$, is assumed to be
    $\gamma=1.3$.  The dashed curve shows the expected redshift
    evolution if the IGM cools adiabatically, $T_{0}\propto
    (1+z)^{2}$, while the dotted curve follows the thermal asymptote,
    $T_{0}\propto (1+z)^{0.53}$.  These curves approximate the maximum
    and minimum rate of cooling expected toward lower redshift.  {\it
      Right:} As for the left panel, except now the near-zone
    temperature constraints are adjusted for the possible systematic
    due to additional Jeans smoothing (see Section~\ref{sec:Jeans} for
    details).  Following \citet{Becker11}, the $z<5$ measurements have
    also been shifted downward by $2000\rm\,K$, reflecting the
    expected systematic shift these authors find at $z \simeq 4$-5
    when including additional Jeans smoothing in their analysis.}
  \label{fig:temp_all}
\end{center}
\end{minipage}

\end{figure*}

The average measurements for all seven quasar are displayed in
Figure~\ref{fig:temp_all}, where they are also compared to the IGM
temperature measurements at $z<5$ presented recently by
\cite{Becker11}.  The left hand panel displays the measurements
obtained using our fiducial thermal history, while the right hand
panel shows our constraints after accounting for additional Jeans
smoothing.  Note that the \cite{Becker11} measurements in the right
panel have been shifted downward by $2000\rm\,K$, reflecting the
approximate expected systematic shift these authors find at $z \simeq
4$-5 when including additional Jeans smoothing in their analysis.

In each panel, we furthermore show two different versions of our
$z\simeq 6$ temperature measurements. The blue circle shows the
temperature constraint obtained directly from the Doppler parameter
CPDF for all seven lines-of-sight.  In addition, the red square shows
the constraint on the {\it initial} temperature of the IGM before the
quasars reionise the \HeII in their vicinity.  Here we have used the
\HeII heating estimates from our radiative transfer simulations to
remove the contribution of the in-situ \HeII photo-heating by the
quasars.  In practice, this is achieved by evaluating
Eq.~(\ref{eq:bayes}) using the initial IGM temperature, $T_{0}^{\rm
  initial}$, in our models rather than the temperature after \HeII
heating by the quasar.  Note that this constraint assumes the average
quasar EUV spectral index is $\alpha_{\rm q}=1.5$; a harder (softer)
spectral index would lower (raise) these constraints.  The advantage
of these measurements is that they attempt to remove the impact of
\HeII heating by the quasar, and should therefore more closely
represent the temperature of the general IGM prior to \HeII
reionisation.  These constraints may therefore be more easily compared
to expectations for the thermal state of the IGM following \HI
reionisation (\citealt{FurlanettoOh09,Cen09}).

We qualitatively explore the implications of these measurements for
reionisation and the IGM thermal history using two simple evolutionary
models for the IGM temperature, shown by the dashed and dotted curves
in Figure~\ref{fig:temp_all}.  The dashed line, which scales as
$T_{0}\propto (1+z)^{2}$, is normalised to match our temperature
constraint including \HeII heating. This curve represents the maximum
possible (adiabatic) cooling rate toward lower redshift in the absence
of any additional heating.  In contrast, the dotted curve scales as
$T_{0}\propto (1+z)^{0.53}$ and is normalised to pass through the
constraint with \HeII heating subtracted.  This curve approximates the
minimum amount of cooling expected if the IGM temperature is
asymptotically approaching the thermal state set by the spectral shape
of the UV background (\citealt{HuiHaiman03}).

Although the absolute values for the temperatures are lower when
including additional Jeans smoothing, the general conclusions we draw
from the relative values of the temperature data remain the same
regardless of the uncertain thermal history at $z>6$.  The results in
Figure~\ref{fig:temp_all} imply that (i) there is an additional and
significant source of heating around the $z\simeq 6$ quasars which is
not yet present in the general IGM at $z\simeq 5$ and (ii) there is
evidence for a constant or gradually increasing temperature in the
general IGM from $z \simeq 6$ to $z\simeq 5$.  The first point
suggests that we may be observing evidence for the earliest stages of
\HeII reionisation in the immediate environment of these high redshift
quasars.  The second point agrees with the suggestion by
\cite{Becker11} that there is a gradual increase in the temperature of
the general IGM at $z<5$ due to the impact of an extended epoch of
\HeII reionisation globally (but see also the suggestion by
\citealt{Chang11} that significant additional heating at $z<5$ may
arise from TeV blazars, assuming the kinetic energy of
ultra-relativistic pairs is converted to thermal energy via plasma beam
instabilities).

\begin{table*}
  \centering
  \caption{The temperature measurements obtained from the Doppler
    parameter CPDF in the near-zones of the seven quasars analysed in
    this study.  From left to right, each column lists the quasar
    name, the redshift range over which the Doppler parameters were
    measured from the spectrum, and the final constraints for our
    fiducial thermal history and for the case of the additional Jeans
    smoothing expected from a very extended period of heating at $z>6$
    (see Section~\ref{sec:Jeans} for details).  The final two rows
    also show the line-of-sight averaged constraints obtained for all
    seven quasars, with the last row giving the temperature
    measurement after subtracting the expected heating from the local
    reionisation of \HeII by the quasar.  In all instances we assume a
    flat prior probability of $3.2 \leq \log (T_{0}/\rm K) \leq 4.8$.}

  \begin{tabular}{c|c|c|c}
   
     \hline
    
     Quasar   & $z$ & log ($T_{0}$/K)  &  log ($T_{0}$/K)   \\
              &      & (Fiducial)       &  (Extra Jeans smoothing)  \\
      \hline
   
    SDSS J1148$+$5251 & $6.38 \pm 0.03$  & $4.10^{+0.10}_{-0.11}$ ($^{+0.19}_{-0.26}$) & $4.02^{+0.12}_{-0.13}$ ($^{+0.23}_{-0.29}$) \\
   \\
    SDSS J1030$+$0524 & $6.27 \pm 0.04$  & $4.10^{+0.07}_{-0.10}$ ($^{+0.13}_{-0.26}$) & $4.01^{+0.10}_{-0.12}$ ($^{+0.20}_{-0.28}$) \\
   \\
    SDSS J1623$+$3112 & $6.21 \pm 0.03$  & $4.26^{+0.10}_{-0.09}$ ($^{+0.19}_{-0.15}$) & $4.18^{+0.12}_{-0.11}$ ($^{+0.23}_{-0.21}$) \\
   \\
    SDSS J0818$+$1722 & $5.98 \pm 0.04$  & $4.24^{+0.10}_{-0.10}$ ($^{+0.19}_{-0.22}$) & $4.16^{+0.12}_{-0.12}$ ($^{+0.25}_{-0.26}$)\\
   \\
    SDSS J1306$+$0356 & $5.98 \pm 0.03$ & $4.24^{+0.11}_{-0.12}$ ($^{+0.20}_{-0.25}$) & $4.15^{+0.13}_{-0.14}$ ($^{+0.25}_{-0.29}$) \\
   \\
    SDSS J0002$+$2550 & $5.79 \pm 0.03$  & $4.33^{+0.09}_{-0.07}$ ($^{+0.27}_{-0.15}$) & $4.25^{+0.12}_{-0.10}$ ($^{+0.28}_{-0.21}$)\\
   \\
    SDSS J0836$+$0054 & $5.78 \pm 0.03$  & $4.26^{+0.09}_{-0.09}$ ($^{+0.18}_{-0.21}$) & $4.18^{+0.11}_{-0.12}$ ($^{+0.23}_{-0.25}$) \\ 
   
    \hline
    All & $6.08\pm 0.33$ & $4.21^{+0.03}_{-0.03}$ ($^{+0.06}_{-0.07}$)  &$4.13^{+0.04}_{-0.04}$ ($^{+0.08}_{-0.08}$) \\
    \\
    All (\HeII heating subtracted)  & $6.08\pm 0.33$ & $3.85^{+0.08}_{-0.08}$ ($^{+0.13}_{-0.17}$)  &$3.77^{+0.08}_{-0.09}$ ($^{+0.14}_{-0.18}$) \\
    \hline
       
\end{tabular}
\label{tab:results}
\end{table*}

\begin{figure}
\begin{center}
  \includegraphics[width=0.45\textwidth]{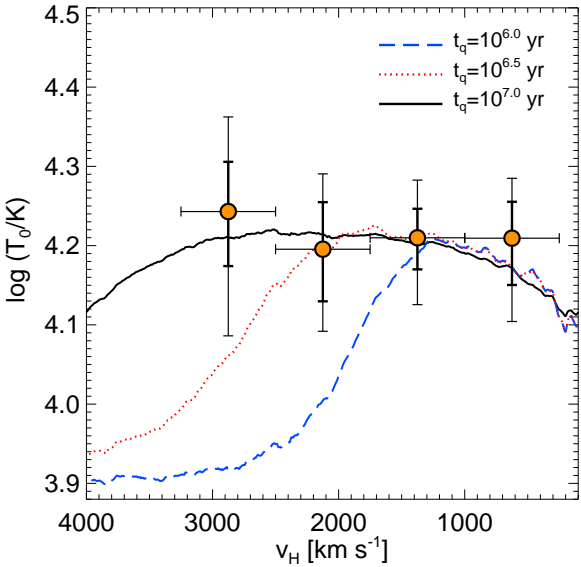}
\vspace{-0.3cm}
\caption{The temperature at mean density as a function of Hubble
  velocity blueward of the quasar \Lya emission lines.  The data
  points are derived from all seven quasars in our sample, where the
  thick (thin) error bars represent the 68 (95) per cent confidence
  intervals.  The solid black, dotted red and dashed blue curves show
  the temperature at mean density predicted by radiative transfer
  simulations constructed from model D for an optically bright phase
  of $t_{\rm q}=10^{7}$, $10^{6.5}$ and $10^{6}$ yr respectively.
  Assuming the quasars reionised \HeII in their environment, a flat
  thermal profile is inconsistent with an optically bright phase of
  duration $t_{\rm q}\leq 10^{6.5}\rm\,yr$.}
\label{fig:temp_prox}
\end{center}
\end{figure}

\subsection{The thermal proximity effect}

Finally, as we suspect these quasars may have recently reionised the
\HeII in their vicinity, it is interesting to also examine the IGM
temperature as a function of distance from the quasars.  We therefore
now consider the possibility of directly observing the line-of-sight
``thermal proximity effect'', which will arise from the elevated
temperatures expected in the \HeIII regions created by these quasars
(\citealt{MiraldaRees94,Theuns02c,Meiksin10}).

Assuming the optically bright phase of a quasar is significantly
shorter than the recombination timescale, $t_{\rm q}\ll t_{\rm
  rec}^{\rm HeIII}$, the typical size of a quasar \HeIII region in
proper Mpc is

\begin{equation} R_{\rm HeIII} \simeq 5.5 {\rm \, Mpc} \left(\frac{{\dot N}_{56}}{3.5}\right)^{1/3}\left(\frac{t_{\rm q}}{10^{7}\rm\,yr}\right)^{1/3}\left(\frac{1+z}{7}\right)^{-1/3}. \label{eq:Rion} \end{equation}

\noindent
Here ${\dot N}_{56}={\dot N}_{\rm HeII}/10^{56}\rm\,s^{-1}$ is the
number of \HeII ionising photons emitted per second, which for the
quasars analysed in this work ranges from ${\dot N}_{56}\simeq
1.7$--$5.3$ assuming an EUV spectral index of $\alpha_{\rm q}=1.5$.  A thermal
proximity effect will thus be detectable only if the quasar \HeIII
regions are on average smaller than the scales over which we make our
temperature measurements ($\sim 5$ proper Mpc). Note also that the
adiabatic cooling timescale, which is the appropriate cooling rate for
gas at mean density expanding with the Hubble flow, is $t_{\rm ad}
\simeq 1/2H(z) = 7.2\times 10^{8}\rm\,yr$ at $z=6$ (cf. the age
of the Universe at $z=6$, $t_{\rm age} \simeq 9.6\times
10^{8}\rm\,yr$). Consequently, the thermal proximity effect should
provide a reasonable constraint on the {\it total} duration of the
optically bright lifetime for these quasars.  We may therefore infer
from Eq.~(\ref{eq:Rion}) that the detection of significantly lower
temperatures at the edge of the quasar near-zones would imply a rather
short optically bright phase on average for these objects, $t_{\rm
  q}<10^{7}\rm\,yr$.

We obtain the radial temperature measurements by splitting the
absorption line fits from all seven quasars into four bins of width
$750\rm\,km\,s^{-1}$ over the range $250\rm\, km\,s^{-1} \leq v_{\rm
  H}\leq 3250\rm\, km\,s^{-1} $.  We then use the Doppler parameter
CPDF in each of the four bins to obtain temperature measurements at
mean density in the usual manner.  Note that due to the small number
of lines in each bin for individual quasars (typically $6-8$), it is
not possible to obtain useful constraints from each quasar
individually.

The results of this procedure are displayed as the data points with
error bars in Figure~\ref{fig:temp_prox}, and are consistent with
constant temperature of $\log (T_{0}/\rm K) \sim 4.2$ within $\sim
3250\rm\, km\,s^{-1}$ of these quasars.  These are compared to a
selection of radial temperature profiles from our line-of-sight
radiative transfer calculations.  The simulations are constructed
using model D, with an initial temperature of $\log (T_{0}/\rm K) \sim
3.9$, and are identical to the models used to construct our synthetic
spectra with the exception that we also now consider two shorter
optically bright phases of $t_{\rm q}=10^{6.5}\rm\,yr$ and $t_{\rm
  q}=10^{6}\rm\,yr$.  The temperature profiles displayed are averaged
over all 100 simulated lines-of-sight and smoothed by a box car window
of width $100\rm\, km\,s^{-1}$ for clarity.  It is evident from this
simple comparison that, under the assumption the quasars have
reionised the \HeII in their vicinity, the data are inconsistent with
an optically bright phase with $t_{\rm q}<10^{6.5}\rm\,yr$.  However,
due to the relatively restricted range probed by the near-zones we are
unable to detect clear evidence for a thermal proximity effect. 

Future progress in this area may be possible at slightly lower
redshift, where line widths can be analysed at larger distances from
the quasar.  However, detecting the thermal proximity effect in quasar
spectra at $z<3$ will likely difficult, since it is expected \HeII
reionisation is largely complete by this time ({\it e.g.}
\citealt{Shull10,Worseck11}).  Analyses at slightly higher redshift,
$z>3.5$, may therefore be best placed for such a study.  On the other
hand, a recent theoretical study by \cite{Meiksin10} found that
peculiar motions and IGM density variations can result in the \Lya
line widths exhibiting very little dependence on the distance from the
quasar, even in the presence of a thermal proximity effect.  Detailed
simulations which correctly model the IGM density and velocity field
will therefore be required to extract temperature measurements from
the data.

%%%%%%%%%%%%%%%%%%%%%%%%%%%%%%%%%%%%%%%%%%%%%%%%%%%%%%%%%%%%%%%%%%%%%	
%%%%%%%%%%%%%%%%%%%%%%%%%% SECTION 6 %%%%%%%%%%%%%%%%%%%%%%%%%%%%%%%%
%%%%%%%%%%%%%%%%%%%%%%%%%%%%%%%%%%%%%%%%%%%%%%%%%%%%%%%%%%%%%%%%%%%%%

\section{Conclusions}

In this work we present improved measurements of the temperature of
the IGM at mean density around $z\simeq 6$ quasars.  We use a sample
of seven high-resolution quasar spectra, combined with detailed
simulations of the thermal state of the inhomogeneous IGM, to improve
upon the first direct measurement of the IGM temperature around
J0818$+1$722 at $z\simeq 6$ presented by B10.  This study therefore
builds upon the work of B10 in three important ways, by using a larger
sample of quasars, an expanded set of numerical simulations for
calibrating the temperature measurements, and a more detailed analysis
of systematic uncertainties.  We find that the most important
systematic is the thermal history of the IGM at $z>6$, which impacts
on our measurements through the uncertain contribution of Jeans
smoothing to the widths of \Lya absorption lines.

The temperature at mean density averaged over all seven lines-of-sight
is $\log(T_{0}/\rm K) = 4.21 \pm ^{0.03}_{0.03}$ ($\pm
^{0.06}_{0.07}$) at 68 (95) per cent confidence.  On comparison to
constraints on the temperature of the general IGM at $z\sim 4.8$ which
are consistent with $\log (T_{0}/\rm K) = 3.9 \pm 0.1$ within
$2\sigma$, these data suggest that there is an additional and
significant source of heating around the $z\simeq 6$ quasars which is
not yet present in the general IGM at $z\simeq 5$.  This is most
likely due to the recent reionisation of \HeII in the vicinity of
these quasars, which is driven by their hard non-thermal ionising
spectra.  The elevated temperatures may therefore represent the first
stages of \HeII reionisation in the most biased locations in the high
redshift Universe.  It is furthermore found that when subtracting the
expected amount of \HeII photo-heating from the quasars, and assuming
a canonical EUV spectral index of $\alpha_{\rm q}=1.5$, the general
IGM temperature is similar to that measured at $z\simeq 4.8$.  This
agrees with the suggestion by \cite{Becker11} that the observed rise
in the IGM temperature at $z\leq 4.4$ is consistent with the onset of
an extended epoch of \HeII reionisation globally.

We also examine the evidence for a line-of-sight thermal proximity
effect around these quasars by analysing the Doppler parameters for
all seven lines-of-sight in radially spaced bins.  We find no clear
evidence for a thermal proximity effect due to an \HeIII region around
the quasar, but note that the limited extent of the near-zone prevents
us from detecting photo-heated \HeIII bubbles larger than $\sim 5$
proper Mpc in size.  Under the assumption that the quasar has
reionised the \HeII in its vicinity, the data are therefore
inconsistent with a short optically bright phase $t_{\rm
  q}<10^{6.5}\rm\,yr$.

Finally, in this work we have not examined the implications of
our temperature measurements for \HI reionisation at $z>6$.  These
high-redshift temperature measurements should still probe the thermal
signature of this landmark event, potentially yielding valuable
insights into the timing and duration of this process ({\it e.g.}
\citealt{Theuns02,HuiHaiman03,FurlanettoOh09,Trac08,Cen09}). We intend
examine this in detail in future work.

\section*{Acknowledgments}

The hydrodynamical simulations used in this work were performed using
the Darwin Supercomputer of the University of Cambridge High
Performance Computing Service (http://www.hpc.cam.ac.uk/), provided by
Dell Inc. using Strategic Research Infrastructure Funding from the
Higher Education Funding Council for England.  JSB acknowledges the
support of an ARC Australian postdoctoral fellowship (DP0984947), and
GDB thanks the Kavli foundation for financial support.

%%%%%%%%%%%%%%%%%%%%%%%%%%%%%%%%%%%%%%%%%%%%%%%%%%%%%%%%%%%%%%%%%%%%%	
%%%%%%%%%%%%%%%%%%%%%%%%% APPENDIX A %%%%%%%%%%%%%%%%%%%%%%%%%%%%%%%%
%%%%%%%%%%%%%%%%%%%%%%%%%%%%%%%%%%%%%%%%%%%%%%%%%%%%%%%%%%%%%%%%%%%%%

\appendix
\section{Tests of the temperature measurement procedure}
\begin{figure*}
\centering
\begin{minipage}{180mm}
\begin{center}
  \psfig{figure=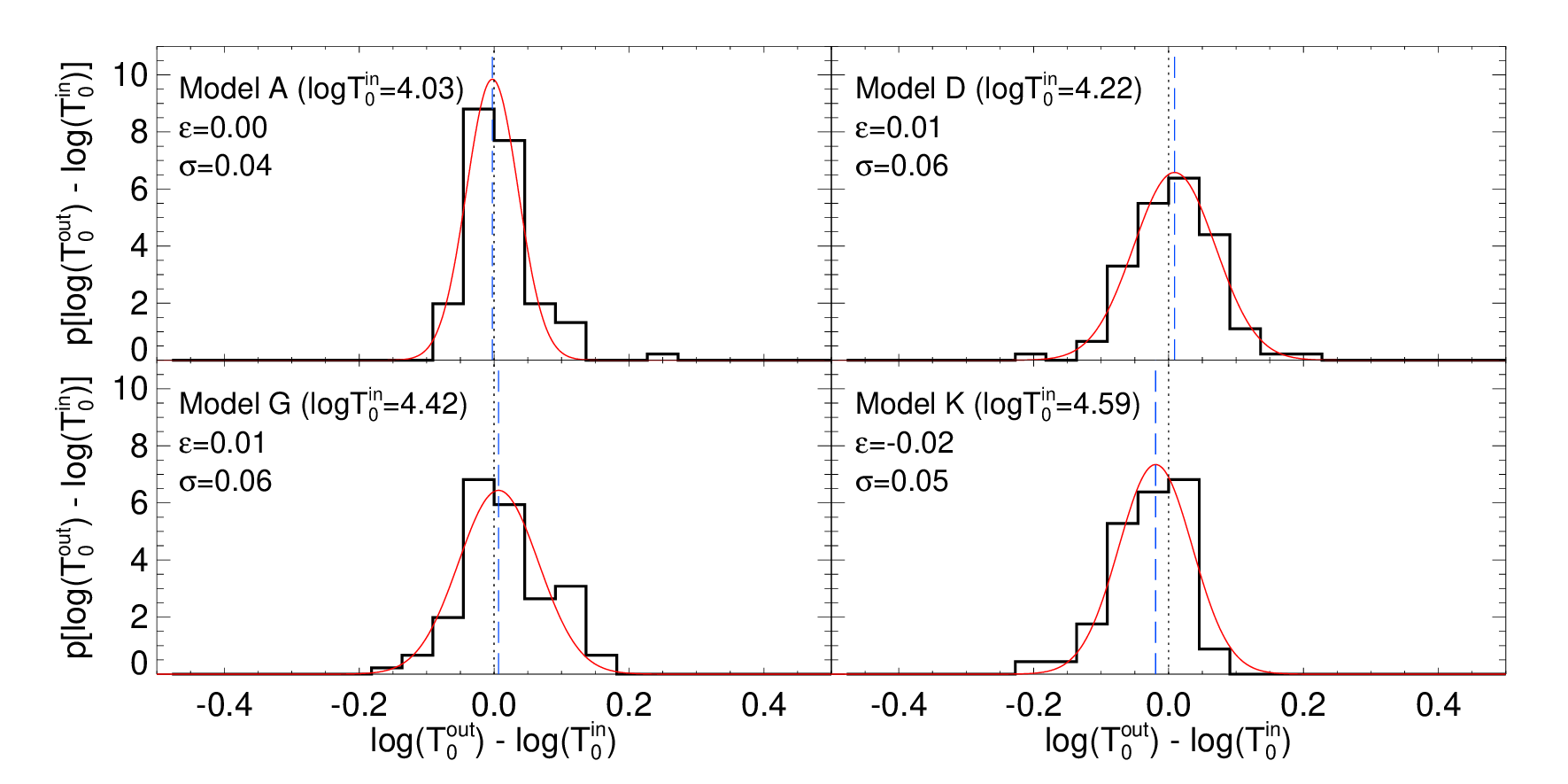,width=0.9\textwidth}
  \vspace{-0.3cm}
  \caption{In each panel the distribution of the difference between
    the temperature at mean density measured from 100 synthetic
    near-zone spectra and the input value, $\log T_{0}^{\rm out}-\log
    T_{0}^{\rm in}$, is displayed by the solid black lines.  The red
    solid curves display the best fit Gaussian to the distribution,
    and the vertical black dotted and dashed blue lines mark the zero
    offset and the mean of the best fit Gaussian, respectively.  {\it
      Upper left:} The $\log T_{0}^{\rm out}-\log T_{0}^{\rm in}$
    distribution for Model A, with a best fit Gaussian with mean
    $\epsilon = 0.00$ dex and standard deviation $\sigma =0.04$ dex.
    {\it Upper right:} Model D, $\epsilon = +0.01$ dex, $\sigma=0.06$
    dex. {\it Lower left:} Model G, $\epsilon = +0.01$ dex,
    $\sigma=0.06$ dex. {\it Lower right:} Model K, $\epsilon = -0.02$
    dex, $\sigma=0.05$ dex.}
  \label{fig:temp_test}
\end{center}
\end{minipage}
\end{figure*}

Tests of the temperature measurement procedure described in
Section~\ref{sec:method} are displayed in Figure~\ref{fig:temp_test}.
The solid black lines show distributions of the difference between the
temperature at mean density measured from 100 synthetic spectra and
the input value, $\log T_{0}^{\rm out}-\log T_{0}^{\rm in}$, for four
of our fiducial models.  The models span a temperature range
consistent with the observational measurements presented in this work.
The distributions may be approximated by a Gaussian with mean
$\epsilon$ and standard deviation $\sigma$, shown by the solid red
curves.  The means are within $\leq 0.02$ dex of the
true input temperature in all cases, indicating our temperature
measurement procedure is reliable to this level in the absence of any
additional systematic uncertainties.  Note, however, the accuracy of
the temperature recovery is slightly poorer for model K, where the
distribution is skewed to lower temperatures.  This will be true in
general for models with progressively higher temperatures; the line
widths scale as the square root of the temperature, $b\propto
T^{1/2}$, and the Doppler parameter CPDF thus has less discriminative
power as the temperature increases.

\end{document}